\documentclass[letterpaper,notitlepage,nofootinbib,superscriptaddress]{revtex4-1}

\usepackage[utf8]{inputenc}
\usepackage{amsmath,amssymb,amsfonts}
\usepackage{graphicx,wrapfig}
\usepackage{dcolumn}
\usepackage{bm}
\usepackage{color}
\usepackage{bbold}
\usepackage{url}
\usepackage{slashed}
\usepackage[svgnames]{xcolor}
\usepackage[english]{babel}
\usepackage{microtype}

\usepackage[colorlinks=true,linkcolor=blue,urlcolor=blue,citecolor=blue]{hyperref}

\usepackage{color}

\definecolor{darkgreen}{rgb}{0,0.65,0}

\newcommand{\be}{\begin{eqnarray}}
\newcommand{\ee}{\end{eqnarray}}
\newcommand{\ba}{\begin{array}}
\newcommand{\ea}{\end{array}}

\newcommand{\Slash}[1]{\ooalign{\hfil/\hfil\crcr$#1$}}

\begin{document}

\title{ Nucleon gravitational form factors from instantons:  forces between quark and gluon subsystems}

\author{Maxim V.~Polyakov}
	\affiliation{Petersburg Nuclear Physics Institute,
		Gatchina, 188300, St.~Petersburg, Russia}
	\affiliation{Institut f\"ur Theoretische Physik II,
		Ruhr-Universit\"at Bochum, D-44780 Bochum, Germany}
\author{Hyeon-Dong Son}
	\affiliation{Institut f\"ur Theoretische Physik II,
		Ruhr-Universit\"at Bochum, D-44780 Bochum, Germany}

\begin{abstract}
Using the instanton picture of the QCD vacuum we compute the nucleon
 $\bar c^Q(t)$ form factor of the quark part of the energy momentum tensor (EMT).
This form factor describes the non-conservation of the quark part of EMT
 and contributes to the quark pressure distribution inside the nucleon. Also it can be interpreted in terms of forces between
 quark and gluon subsystems inside the nucleon.
 We show that this form factor is parametrically small
in the instanton packing fraction.
 Numerically we obtain for the nucleon EMT a small value of
 $\bar c^Q(0)\simeq 1.4\cdot 10^{-2}$ at the low normalisation point of $\sim 0.4$~GeV$^2$.
{This smallness
implies interesting physics picture --
the forces between quark and gluon mechanical subsystems   are smaller than the forces inside each subsystem.
The forces from side of gluon subsystem squeeze the quark subsystem -- they are compression forces. Additionally, the smallness of $\bar c^Q(t)$ might
justify  Teryaev's equipartition conjecture. }
We estimate
that the contribution of $\bar c^Q (t)$ to the pressure distribution inside the nucleon  is in the range of $1 -20$\%  relative to the
contribution of the quark $D$-term.
\end{abstract}


\maketitle

\section*{\normalsize \bf Introduction}
The hadron form factors of energy momentum tensor (EMT) were introduced in 1960's in Refs.~\cite{Kobzarev:1962wt,Pagels:1966zza}
to study the behaviour of hadrons in curved space-time and to obtain the basic mechanical properties of them.
Nowadays the interest to EMT form factors increased as they can be, in principle, accessed in hard exclusive processes without invoking
very weak gravitational forces and in this way to study in details the mechanical properties of the hadrons.

The symmetric QCD energy-momentum tensor operators for quark and gluon can be
obtained by varying the QCD action in respect to the metric of curved space-time, it has the
following form
\begin{align}\label{eq:qcd_EMT}
	T^{\mu\nu}_q &= \frac{1}{4}\overline{\psi}_q\biggl(
	-i\overset{ \leftarrow}{\cal D}{ }^\mu\gamma^\nu
	-i\overset{ \leftarrow}{\cal D}{ }^\nu\gamma^\mu
	+i\overset{\rightarrow}{\cal D}{ }^\mu\gamma^\nu
	+i\overset{\rightarrow}{\cal D}{ }^\nu\gamma^\mu\biggr)\psi_q
	-g^{\mu\nu}\overline{\psi}_q\biggl(
	-\frac{i}{2}\,\overset{ \leftarrow}{\Slash{\cal D}}{ }
	+\frac{i}{2}\,\overset{\rightarrow}{\Slash{\cal D}}{ }
	{\,-\,m_q}\biggr)\psi_q, \\
	T^{\mu\nu}_g &= F^{a,\mu\eta}\,{F^{a,}}_{\eta}{ }^\nu+\frac14\,g^{\mu\nu}
	F^{a,\kappa\eta}\,{F^{a,}}_{\kappa\eta}.
\end{align}
Here $\overset{\rightarrow}{\cal D}_\mu=\overset{\rightarrow}\partial_\mu-ig\,t^aA_\mu^a$
and $\overset{\leftarrow}{\cal D}_\mu = \overset{\leftarrow}\partial_\mu+ig\,t^aA_\mu^a$
with arrows indicating which fields are differentiated,
$F^a_{\mu\nu}=\partial_\mu A^a_\nu-\partial_\nu A^a_\mu+g\,f^{abc}A^b_\mu A^c_\nu$
and the SU(3) color group generators satisfy the algebra
$[t^a,t^b]=i\,f^{abc}t^c$ and are normalized as
${\rm tr}\,(t^at^b)=\frac12\,\delta^{ab}$.
The total EMT is conserved
\be\label{Eq:EMT-cons}
	\partial^\mu T_{\mu\nu} = 0, \quad \quad
	 T_{\mu\nu} = \sum_q T_{\mu\nu}^q+ T_{\mu\nu}^g \; .
\ee
The nucleon matrix element of individual pieces of EMT operator can be parameterized
as the following expression,
\begin{align}
    & \langle p^\prime,s^\prime|  T_{\mu\nu}^a(x) |p,s\rangle \cr
    &= \bar u^\prime\biggl[
      A^a(t)\,\frac{P_\mu P_\nu}{M_N}
    + J^a(t)\ \frac{i\,P_{\{\mu}\sigma_{\nu\}\rho}\Delta^\rho}{2M_N}
    + D^a(t)\,\frac{\Delta_\mu\Delta_\nu-g_{\mu\nu}\Delta^2}{4M_N}
    +{M_N}\,{\bar c}^a(t)g_{\mu\nu} \biggr]u\,\,e^{i(p^\prime-p)x}\cr
    \label{eq:EMT-FFs-spin-1/2}
\end{align}
We introduced the notation $a_{\{\mu} b_{\nu\}}=a_\mu b_\nu + a_\nu b_\mu$, as well as $P=(p'+p)/2$, $\Delta = p'-p$.
The spinors satisfy the normalization condition,
$\bar u(p,s)\, u(p,s) =2 M_N$ where $M_N$ is the nucleon mass.
Due to EMT conservation, Eq.~(\ref{Eq:EMT-cons}), the constraint
$\sum_a\bar{c}^a(t)=0$ holds

The physics interpretation of the EMT form factors, their calculation in various models, and extraction from experimental data were
extensively discussed in recent review \cite{Polyakov:2018zvc}.  Here we concentrate on the form factor ${\bar c}^Q(t)=\sum_{a=u,d,s, \ldots}\bar{c}^a(t)$,
which describes the non-conservation of EMT for individual quark and gluon pieces.
 This form factor is important to determine the pressure forces distribution in the nucleon individually for
quarks and gluons, and to study the forces between quark and gluon subsystems in the nucleon.
The form factor $\bar c^Q(t)$ is the least studied,
we are aware only about the calculation of $\bar c^Q(t)$ in the bag model with the result of $\bar c^Q(0)\simeq -1/4$ \cite{Ji:1997gm}.
The value resulted from the relation $\bar c^Q(0)=-A^Q(0)/4$ in the bag model \cite{Ji:1997gm}, however the authors of \cite{Ji:1997gm} stressed that  this relation is not true in
QCD because the renormalised quark part of the energy-momentum tensor has a trace anomaly. The relations of $\bar c^Q(t)$ to
twist-4 generalised parton distributions (GPDs) were derived in Refs.~\cite{Leader:2012ar,Leader:2013jra,Tanaka:2018wea}.

The individual quark and gluon form factors
$A^a(t)$,  $J^a(t)$, $D^a(t)$, ${\bar c}^a(t)$ depend
on the renormalization scale which we do not indicate for brevity.
The scale dependence of the quark form factor  $\bar{c}^Q(t)$ has (at one loop level) the following form:

\be
\label{eq:evolution}
{\bar c}^Q(t)|_{\mu}= {\bar c}^Q(t)|_{\mu_0} \left[\frac{\alpha_s(\mu)}{\alpha_s(\mu_0)}\right]^{\gamma/b},
\ee
where $b=\frac{11}{3} N_c-\frac 23 N_f$ is the one-loop coefficient of the QCD $\beta$-function and $\gamma=\frac{4(N_c^2-1)}{3 N_c}+\frac 23 N_f$ is the corresponding anomalous dimension.
With the increasing of the normalisation point the form factor ${\bar c}^Q(t)$ goes logarithmically to zero. Important question is: what is the size of ${\bar c}^Q(t)$
at a low normalisation point  -- initial condition for the QCD evolution? Our aim here is to answer this question.

Using the QCD equation of motion one can easily relate (see e.g. \cite{Kolesnichenko:1984,Braun:2004vf,Tanaka:2018wea})
the quark form factor $\bar{c}^Q(t)$
to the matrix element of the flavour singlet twist-4 quark-gluon operator\:

\begin{align}\label{eq:tanaka-cbar-operator-relation}
\Delta^\beta \bar u(p',s')u(p,s) M_N \bar c^Q(t) =
\langle p',s' | ig \bar{\psi}  F^{\beta\alpha} \gamma_\alpha \psi | p, s \rangle.
\end{align}
Analogously one can obtain the expression for the gluon form factor $\bar c^g(t)$ (see e.g. \S 33 of \cite{Landau:1987gn}, \cite{ANV}, or \cite{Tanaka:2018wea}):

\be
\label{eq:tanaka-cbar-operator-relation-gluon}
\Delta^\beta \bar u(p',s')u(p,s) M_N \bar c^g(t) =
\langle p',s' | \frac 12\  {\rm tr}\left (F^{\beta \alpha}\left[ i{\cal D}^\sigma, F_{\sigma\alpha}\right] \right)| p, s \rangle.
\ee
It is obvious from two expressions above that $\bar c^Q(t)+\bar c^q(t)=0$ because of QCD equation of motion $\left[{\cal D}^\sigma, F_{\sigma\alpha}\right] =j_\alpha^a t^a$, with
$j_\alpha^a =-g\ \bar \psi \gamma_\alpha  t^a \psi$.

In present paper we estimate the matrix element  (\ref{eq:tanaka-cbar-operator-relation}) in the instanton picture of the QCD vacuum using the technique of effective operators
developed in Ref.~\cite{Diakonov:1995qy}. This technique was applied to matrix elements of higher twist operators in
\cite{Balla:1997hf,Dressler:1999hc,Dressler:1999zi,Lee:2001ug,Kiptily:2002nx}.
Equally well we can compute with this technique  $\bar c^g(t)$ using (\ref{eq:tanaka-cbar-operator-relation-gluon}), but this is not necessary as in Ref.~\cite{Balla:1997hf}
it was demonstrated that the QCD equations of motion are satisfied by the effective operators in the instanton vacuum.

The approach is based on the expansion in the small parameter intrinsic to the instanton vacuum \cite{Shuryak:1981ff,Diakonov:1983hh} -- instanton packing fraction
$\bar{\rho}/\bar{R}\simeq 1/3$, where $\bar{\rho}$ is the average size of the instanton and $\bar{R}$ the average distance between them.
The relevance of this vacuum parameter to physics of higher twist matrix elements was realised in an interesting way. Back in year 1997 in Ref.~\cite{Balla:1997hf}
it was shown that the nucleon matrix element of the twist-3 quark-gluon operator $d^{(2)}$  (the third Mellin moment of transverse polarized
structure function $g_T(x)$) is suppressed by the instanton packing fraction as $\sim \bar{\rho}^4/\bar{R}^4$, that
leads to the prediction of the numerical value of $d^{(2)}$ which was about several tens (!) times smaller than other predictions
at that time (of QCD sum rules, bag model and lattice, see Table~2 of Ref.~\cite{Balla:1997hf}), the experimental data were not conclusive. With advent of new generation of experiments (see e.g. Ref.~\cite{Anthony:2002hy})
the prediction of \cite{Balla:1997hf} of the suppression of $d^{(2)}$ due to instanton dilutness was confirmed. In our view, the experimentally observed smallness of the third Mellin
of $g_T(x)$ is the strong support of the instanton picture of the QCD vacuum.

We show below that the nucleon EMT form factor $\bar c^Q(t)$ is suppressed in the instanton dilutness as $\sim \bar{\rho}^4/\bar{R}^4$ and numerically small.

\noindent
\section*{\normalsize \bf Effective quark-gluon operators in the Instanton vacuum}

In this section we give only the basics of instanton picture of the QCD vacuum and of the technique of the effective operators,
details can be found e.g. in Refs.~\cite{Diakonov:1983hh,Diakonov:1985eg,Diakonov:1995qy,Balla:1997hf}.
In the instanton picture the QCD partition function can be approximated in terms of the
dilute instanton configuration in the large $N_c$.
The
average (anti-)instanton size $\bar{\rho} \sim 0.3\  \mathrm{fm}$ and
inter-instanton distance $\bar R \sim 1\ \mathrm{fm}$ are
  obtained  phenomenologically in Refs.~\cite{Shuryak:1981ff} and by a variational method in Ref.~\cite{Diakonov:1983hh}. In this picture the QCD vacuum is described as
  {\it dilute} instanton liquid with intrinsic small parameter -- the instanton packing fraction $\bar{\rho}/\bar{R}\simeq 1/3$.

 This picture naturally leads to the spontaneous
 chiral symmetry breaking and describes well
 fundamental low-energy properties of QCD \cite{Diakonov:1985eg}.
 The effective low-energy theory one derives from the instanton vacuum is formulated in terms of degrees of freedom which are pions (Goldstone bosons) and massive constituent
 quarks (see detailed lectures of D.~Diakonov on that in Refs.~\cite{Diakonov:1995ea}). It is described by the effective action:

 \begin{align}\label{eq:quark_effective_action_nf1}
 S_{\mathrm{eff}}
 = \int d^4 x  \bar{\psi} (x)
 ~\left[i \gamma^\mu \partial_\mu -  M F(\overset{\leftarrow}\partial) U^{\gamma_5}(x) F(\overset{\rightarrow}\partial)\right] ~\psi(x).
 \end{align}
 Here, $M$ is the dynamical quark mass generated by the spontaneous breaking of chiral
symmetry; parametrically it is small and is of order:

 \be
 \label{eq:Msmallness}
 M \bar{\rho} \sim \left( \frac{\bar{\rho}}{\bar{R}}\right)^2.
 \ee
 The momentum dependence of the quark mass is obtained
from the Fourier transformation of the instanton zero-mode,
\begin{align}\label{eq:F_zeromode}
F(k) = 2 t \left[ I_0(t)K_1(t)-I_1(t)K_0(t)-\frac{1}{t}I_1(t)K_1(t) \right]
, \qquad t = k \bar{\rho}/2,
\end{align}
normalized as $F(0)=1$ and where $I_n(t)$, $K_n(t)$ are modified Bessel functions
 of the first and second kind, respectively.
The nonlinear unitary matrix field $U^{\gamma_5}(x)$ describes the Goldstone degrees of freedom:

\be
\label{eq:Ugamma5def}
U^{\gamma_5}(x)=U(x) \frac{1+\gamma_5}{2}+U^\dagger(x) \frac{1-\gamma_5}{2}, \quad U(x)=e^{i \pi^a(x) \tau^a/F_\pi}.
\ee
In Refs.~\cite{Diakonov:1995qy,Balla:1997hf} a method to express quark-gluon QCD operators in terms of effective degrees of freedom was developed.
Using this method we can express the operator of our prime interest here:

\be
\label{eq:OpQCD}
O^\beta(x)=ig \bar{\psi}  F^{\beta\alpha} \gamma_\alpha \psi,
\ee
as the following effective operator\footnote{We present the operator rotated to the Euclidean space (Wick rotated). }:

\begin{align}\label{eq:gluon_operator-2}
O^{\rm eff}_{\beta}(x)
&=\frac{i M}{N_c}
\int d^4z \mathcal{F}_{\mu\nu\beta\alpha}(x-z)
\left[-i \psi^\dagger(x) t^a
 \gamma_\alpha \psi(x)\right]
\left[
 \psi^\dagger(z) F(\overset{\leftarrow}{\partial})
  t^a \sigma_{\mu\nu} U^{\gamma_5}(z)
  F(\overset{\rightarrow}{\partial}) \psi(z)
\right]
\end{align}
with
\begin{align}\label{eq:F}
\mathcal{F}_{\mu\nu\beta\alpha}(x) =
\frac{8 \bar{\rho}^2}{(x^2 + \bar{\rho}^2 )^2}
\left(
 \frac{x_\mu x_\beta}{x^2}\delta_{\alpha \nu}
+\frac{x_\nu x_\alpha}{x^2}\delta_{\mu \beta}
-\frac12 \delta_{\mu\beta}\delta_{\alpha \nu}
\right).
\end{align}
Now the matrix elements of the QCD operator (\ref{eq:OpQCD}) can be approximated by the matrix element of the
effective operator  (\ref{eq:gluon_operator-2}) computed in the effective theory (\ref{eq:quark_effective_action_nf1}).
The result of such calculation corresponds to the QCD normalisation point of $\mu^2\sim 1/\bar{\rho}^2\sim 0.4$~GeV$^2$,
see Refs.~\cite{Diakonov:1983hh,Diakonov:1995qy}.
\noindent
\section*{\normalsize \bf Form factor $\mathbf{\bar{c}_{quark}(t)}$ of the constituent quark}
In order to find the parametric dependence of the nucleon form factor $\bar c^Q(t)$ on the instanton packing fraction $\bar{\rho}^2/\bar R^2$
it is enough to compute this form factor for a single constituent quark, which we define
analogously to Eq. (\ref{eq:tanaka-cbar-operator-relation}):

\begin{align}\label{eq:tanaka-cbar-operator-relation-quark}
\Delta^\mu \bar u(p',s')u(p,s) M \bar c_{\mathrm{quark}}(t) =
\langle p',s' | ig \bar{\psi} F^{\mu\nu} \gamma_\nu \psi | p, s \rangle.
\end{align}
As we discussed above, we reduce the calculation of single quark matrix element of effective operator (\ref{eq:gluon_operator-2}) in the
effective theory (\ref{eq:quark_effective_action_nf1}).
The result of simple calculation in the leading order of the $1/N_c$ expansion is:
\begin{align}\label{eq:mt_element}
\langle p', s' | O^{\rm eff}_\beta(x)& | p, s\rangle
= \frac{M}{2}\int \frac{d^4k}{(2\pi)^4}
\mathcal{F}_{\mu\nu\beta\alpha}(k) \cr
	&\left\{\frac{F(p)F(p-k)}{(p-k)^2+M^2F^4(p-k)}{u^\dagger}(p',s') \gamma_\alpha
	(\Slash{p}-\Slash{k} +i MF^2(p-k) )\sigma_{\mu\nu}
	 u(p,s) \right.\cr
	&\left. +\frac{F(p')F(p'+k)}{(p'+k)^2+M^2F^4(p'+k)}
	{u^\dagger}(p',s')\sigma_{\mu\nu}
	(\Slash{p'}+\Slash{k} +i MF^2(p'+k) )\gamma_\alpha u(p,s) \right\}. \cr
\end{align}
Here, $\mathcal{F}(k)$ is the Fourier transform of the field strength on the instanton:

\begin{align}\label{eq:F_momentum}
\mathcal{F}_{\mu\nu\alpha\beta}(k) =
\bar{\rho}^2 \mathcal{G}(k) \left(
  \frac{k_\mu k_\beta}{k^2}\delta_{\alpha \nu}
+ \frac{k_\alpha k_\nu}{k^2}\delta_{\beta \mu}
- \frac{1}{2}\delta_{\beta \mu}\delta_{\alpha \nu} \right),
\end{align}
with
\begin{align}\label{eq:G}
\mathcal{G}(k) = 32 \pi^2\left[
\left(\frac{1}{2}+\frac{4}{t^2}\right)K_0(t)
+\left( \frac{2}{t}+\frac{8}{t^3} \right)K_1(t) - \frac{8}{t^4}
\right], \quad t = k \bar{\rho}.
\end{align}

Further using the equation of motion for the Dirac spinors,
 the form factor is readily obtained from its definition,
  Eq. (\ref{eq:tanaka-cbar-operator-relation-quark}) .
\begin{align}\label{eq:formfactor}
\bar{c}_{\mathrm{quark}} = \frac{M^2}{p^2} I(p^2),
\end{align}
where
\begin{align}\label{eq:integral_I}
I(p) = \bar{\rho}^2 \int \frac{d^4k}{(2\pi)^4}
\mathcal{G}(k) \frac{F(p)F(p-k)^4}{(p-k)^2+M^2F(p-k)^4}\frac{1}{3}
\left( 1 - 4 \frac{(k\cdot p)^2}{k^2 p^2}\right).
\end{align}

In the low energy limit $p\ll 1/\bar \rho$
 the integral $I(p)$ can be computed with the logarithmic accuracy as:
\begin{align}\label{eq:limit1}
& I(p) \sim \frac{1}{12}p^2
\bar{\rho}^2 \ln \left(\frac{1}{p^2 \bar{\rho}^2}\right)
, \qquad \mathrm{for}\;\; M \ll p \ll 1/\bar{\rho}, \\
& \label{eq:limit2}
I(p) \sim \frac{1}{12}p^2
\bar{\rho}^2 \ln \left(\frac{1}{M^2 \bar{\rho}^2}\right)
, \qquad \mathrm{for}\;\; p \ll M \ll 1/\bar{\rho}.
\end{align}
This immediately tells us the parametric dependence of $\bar{c}_{\mathrm{quark}}$ on the packing fraction:

\be
\bar{c}_{\mathrm{quark}}\sim \frac{1}{6}\ (M\bar\rho)^2 \ln\left(\frac{1}{M\bar\rho}\right).
\ee
Given the smallness of the constituent quark mass $M$ in the packing fraction (\ref{eq:Msmallness}) we obtain that $\bar{c}_{\mathrm{quark}}\sim \bar\rho^4/\bar R^4 \ln\left(\bar R/\bar\rho\right)$
is strongly suppressed by the instanton packing fraction.
Numerically for $M\bar{\rho}=0.58$, which is a phenomenological value \cite{Diakonov:1985eg}, we obtain from Eq.~(\ref{eq:formfactor}):
\begin{align}\label{eq:qff_numerical}
\bar{c}_{\mathrm{quark}}
 \simeq 1.4 \cdot 10^{-2}.
\end{align}
Remarkably small number!

\noindent
\section*{\normalsize \bf Estimate of the nucleon form factor $\mathbf{\bar{c}^Q(t)}$}

The nucleon in the effective theory (\ref{eq:quark_effective_action_nf1}) emerges as a soliton of the chiral field \cite{Diakonov:1986yh,Diakonov:1987ty}.
The nucleon matrix element of the effective operator (\ref{eq:gluon_operator-2}) can be computed as a double sum over quark orbitals in the background
chiral soliton field, see examples of such calculations in Section~4.2 of Ref.~\cite{Balla:1997hf}.

Here, for a quick estimate of $\bar c^Q(t)$, we do further approximation. The exact calculation of the sums over quark orbitals will be presented elsewhere.
The effective operator (\ref{eq:gluon_operator-2}) approximate the QCD operator for the momenta $p\le 1/\bar\rho$. If we now integrate out the momenta
of order $M\le p\le 1/\bar\rho$\footnote{Contracting two quark fields in (\ref{eq:gluon_operator-2}) and using  (\ref{eq:formfactor},\ref{eq:limit2}).} we obtain the effective operator
(\ref{eq:gluon_operator-2}) in simplified form:

\be
\label{eq:OeffLE}
O_\beta^{\rm eff}(x)\simeq -i \bar c_{\rm quark}\ \partial_\beta \left(\bar\psi(x) M U^{\gamma_5}(x)\psi(x)\right),
\ee
where $\bar c_{\rm quark}$ is given by (\ref{eq:formfactor}) and numerically by (\ref{eq:qff_numerical}). Although the above effective
operator valid for very small quark momenta $p\sim M$ it can used for the order of magnitude estimate of the nucleon form factor.

Using the simplified effective operator (\ref{eq:OeffLE}) we can compute the nucleon $\bar c^Q(t)$ as:

\begin{align}\label{eq:mt_effective_nucleon}
  M_N \bar{c}^Q_N(t) \bar{u}(p',s') u(p,s)
 =
 \langle p',s' |
 \bar{c}_{\mathrm{quark}} \bar{\psi}M U^{\gamma_5} \psi
 |p,s \rangle .
 \end{align}
Remarkably that  the forward nucleon matrix element of the operator  $\bar{\psi}M U^{\gamma_5} \psi$ can be computed exactly
in the effective theory as\footnote{It can be computed using the virial theorem for the effective theory (\ref{eq:quark_effective_action_nf1}), see Section~3.5 of Ref.~\cite{Diakonov:1996sr}}:

\be
\langle p| \bar{\psi}M U^{\gamma_5} \psi|p\rangle=2 M_N^2.
\ee
 This result immediately implies that:
 \begin{align}\label{eq:ff_effective_nucleon}
\bar{c}^Q(0) = \bar{c}_{\mathrm{quark}}\simeq 1.4\cdot 10^{-2}.
\end{align}
The $t$-dependence of the matrix element in (\ref{eq:mt_effective_nucleon}) is similar to that of the EMT form factor $A(t)$.
 The later was computed in the chiral quark-soliton model(ChQSM) and presented in the form of the dipole Ansatz in Ref.~\cite{Goeke:2007fp}.
 Assuming the form factor $c^Q(t)$ has the same t-dependence as A(t) in \cite{Goeke:2007fp}, we present our result as follows:

 \begin{align}\label{eq:ff_dipole}
\bar c^Q (t)= \frac{\bar{c}_{\rm quark}}{\left(1-t/\Lambda^2\right)^2}.
\end{align}

At the zero momentum transfer, $\bar{c}^Q(0) = \bar{c}_{\mathrm{quark}}\simeq 1.4\cdot 10^{-2}$ is given in Eq. (\ref{eq:ff_effective_nucleon}) and the pole-mass $\Lambda \sim 0.9$~GeV$^2$ is taken from Ref.~\cite{Goeke:2007fp}.
The equation refers to the low normalisation point of $\mu^2\sim 1/\bar\rho^2 \sim 0.4$~GeV$^2$ and the values for higher normalisation points can be obtained with help of Eq.~(\ref{eq:evolution}).

In the large $N_c$ limit we obtain that  $\bar c^Q (t)\sim N_c^0$, it is also easy to see that the flavour non-singlet form factor $\bar c^{u-d}(t) \sim 1/N_c$ in this limit.

 The Eq.~(\ref{eq:ff_dipole}) is the main result of our paper -- the nucleon form factor $\bar c^Q(t)$ is very small even at the low QCD normalisation point of $\mu^2\sim 0.4$~GeV$^2$.
 Our result is about 20 times smaller and has an opposite sign
than the result of the bag model calculation in Ref.~\cite{Ji:1997gm}.
The physics reason for such suppression is well understood -- it is related to the diluteness of the instantons in the QCD vacuum.  We did rough estimate of the nucleon
 $\bar c^Q(t)$, however it is obvious that the more elaborated calculations will change the numerical factor of order one in Eq.~(\ref{eq:ff_effective_nucleon}) but
 not the smallness of the form factor $\sim 10^{-2}$. From this smallness we can obtain interesting physics picture of forces inside the nucleon.

\section*{Pressure distribution in the nucleon and forces between quark and gluon subsystems}

The distribution of forces inside the nucleon can be accessed through the static stress tensor $T_{ij}^a(\bm{r})$ \cite{Polyakov:2002yz}
which is defined in the Breit frame as:

\be\label{EQ:staticEMT}
	T^a_{ij}(\bm{r})=\int \frac{d^3 \bm{\Delta}}{(2\pi)^3 2E}\
	e^{{-i} \bm{r\Delta}} \langle p'| {T}^a_{ij}(0)|p\rangle.
\ee
The static stress tensor can be decomposed in a
traceless part associated with shear forces $s^a(r)$ and a trace part
associated with the pressure $p^a(r)$, 

\begin{align}\label{Eq:stress-tensor-p-s}
	T_{ij}^a(\bm{r}) = \biggl(
	\frac{r_ir_j}{r^2}-\frac13\,\delta_{ij}\biggr) s^a(r)
	+ \delta_{ij}\,p^a(r)\,.
\end{align}
The pressure and shear force distributions of the nucleon can be computed in terms of EMT form factors (\ref{eq:EMT-FFs-spin-1/2}) as
\cite{Polyakov:2018zvc}:
\begin{align}\label{eq:pressure_distribution}
s^a(r)&= -\frac{1}{4 M_N}\ r \frac{d}{dr} \frac{1}{r} \frac{d}{dr}
	{\widetilde{D}^a(r)}, \\
p^a(r)&=\frac{1}{6 M_N} \frac{1}{r^2}\frac{d}{dr} r^2\frac{d}{dr}
	{\widetilde{D^a}(r)}
	- M_N \int {\frac{d^3\Delta}{(2\pi)^3}}\ e^{{-i} \bm{\Delta r}}\
	 \bar c^a(-\bm{\Delta}^2).
\end{align}
with ${\widetilde{D}^a(r)=}
	\int {\frac{d^3\Delta}{(2\pi)^3}}\ e^{{-i} \bm{\Delta r}}\ D^a(-\bm{\Delta}^2)$.
	Note that the form factor $\bar c^a(t)$ contributes only to the pressure inside the nucleon.

For the total (quarks+gluons) stress tensor $T_{ij}=T_{ij}^Q+T_{ij}^g$ the stability (equilibrium) condition is
$
\frac{\partial T_{ij}(\bm{r})}{\partial r_j} =0,
$
for the quark part of the stress tensor the equation reads:

\be
\label{eq:equilibrium}
\frac{\partial T_{ij}^Q(\bm{r})}{\partial r_j} +f_i(\bm{r})=0.
\ee
This equation can be interpreted (see e.g \S 2 of \cite{LLv7}) as equilibrium
equation for quark internal stress and external force (per unit of the volume) $f_i(\bm{r})$ from the side of the gluons.
 This gluon force can be
computed in terms of EMT form factor $\bar c^Q(t)$ as:

\be
\label{eq:specificforce}
f_i(\bm{r})=M_N \frac{\partial}{\partial r_i} \int {\frac{d^3\Delta}{(2\pi)^3}}\ e^{{-i} \bm{\Delta r}}\
	 \bar c^Q(-\bm{\Delta}^2)
\ee
Due to spherical symmetry this force (per unit of volume) is directed along unit vector $n_i=r_i/r$. For the case of monotonically decreasing  with distance
Fourier transform of $\bar c^Q(t)$ (in practice for  $\bar c^Q(0)>0$) the corresponding force (\ref{eq:specificforce}) is directed towards the nucleon centre,
therefore we call it squeezing (compression) force. For opposite
sign the corresponding force is stretching. The results of previous sections imply that the gluon forces squeeze (compress) the quark subsystem.

Integrating Eq.~(\ref{eq:specificforce}) over some volume
we obtain the force acting on this volume from side of gluons. Taking a spherical ball of the radius $R$ we can easily obtain that the total gluon force which
squeezes  (compresses) the quarks has the value:
\be
F(R)=8\pi M_N \int_0^R dr\ r \int {\frac{d^3\Delta}{(2\pi)^3}}\  \left( e^{{-i} \bm{\Delta r}}-e^{{-i} \bm{\Delta n} R}\right)\
	 \bar c^Q(-\bm{\Delta}^2)
\ee
The total squeezing gluon force acting on quarks in the nucleon is equal to $F_{\rm total}=F(\infty)$:

\be
F_{\rm total}= \frac{2 M_N}{ \pi} \int_{-\infty}^0 \frac{dt}{\sqrt{-t}}\  \bar c^Q(t).
\ee
The estimates of the nucleon $\bar c^Q (t)$ in previous sections can be parametrised by a simple dipole Ansatz (\ref{eq:ff_dipole}).
With this Ansatz we obtain that the total squeezing (compression) gluon force acting on the quark subsystem in the nucleon is:

\be
F_{\rm total}=\bar{c}_{\rm quark}\ M_N\Lambda\simeq 5.9\cdot 10^{-2} \frac{\rm GeV}{\rm fm}.
\ee
This force can be compared with typical size of forces inside the quark subsystem. The latter in the nucleon are of order $\sim 0.2$~GeV/fm \cite{Polyakov:2018zvc},
i.e. intersystems force is about 3 times smaller. Also this force is about 15 times smaller than the confinement force $\sim 1$~GeV/fm commonly associated with the  string tension.
So, we have an interesting physics picture -- the interaction between quark and gluon subsystems   weaker than the
interaction inside the quark subsystem.

From Eq.~(\ref{eq:pressure_distribution}) we see that the form factor $\bar c^Q (t)$ contributes to the pressure distribution in the nucleon. With our parametrization (\ref{eq:ff_dipole})
one can easily obtain that the contribution of $\bar c^Q (t)$  to quark pressure at the origin (the highest value) is of order $\sim -0.045$~GeV/fm$^3$\footnote{We note that the contribution
of $\bar c^Q(t)$ to the pressure ($\Delta p(r)$) is very sensitive to the choice of the parameter $\Lambda$ in the Ansatz (\ref{eq:ff_dipole}) ($\Delta p(0)\sim M_N \Lambda^3$). This leads to large uncertainties
due to poor knowledge of $\Lambda$}.
Using the calculations of the total pressure in \cite{Goeke:2007fp} and its phenomenological extraction from DVCS data in \cite{Nature} we estimated
that the contribution of $\bar c^Q (t)$ to the pressure distribution  inside the nucleon  is in the range of $1 -20$\% of the
contribution of the $D$-term.
\noindent
\section*{\normalsize \bf Conclusion and discussion }

The quark and gluon pieces of the energy-momentum tensor (EMT) are not conserved individually, only their sum is conserved. The form factor $\bar c^Q(t)$
describes the size of the non-conservation of the quark part of EMT. The corresponding non-conservation relation can be interpreted as equilibrium
equation for quark internal stress and external forces from the side of the gluons, see Eq.~(\ref{eq:equilibrium}).
In that equation gluon force (per unit of the volume) $f_i(\bm{r})$
computed in terms of EMT form factor $\bar c^Q(t)$ (\ref{eq:specificforce}). In this way this EMT form factor characterises the strength of the interaction between
quark and gluon subsystems.

We computed the nucleon form factor $\bar c^Q(t)$ in the instanton picture of the QCD vacuum. It turned out that this form factor is parametrically small
in the  instanton packing fraction $\bar{c}^Q(t)\sim \bar\rho^4/\bar R^4 \ln\left(\bar R/\bar\rho\right)$. This parametric suppression is reflected in small
numerical value of $\bar c^Q(0)\simeq 1.4\cdot 10^{-2}$ we obtained at the low normallisation point of  $\mu^2\sim 1/\bar\rho^2 \sim 0.4$~GeV$^2$.
For higher normalisation points this form factor logarithmically goes to zero according to the evolution equation (\ref{eq:evolution}).

 The numerical value for $\bar c^Q(t)$ we obtained is about hundred times smaller than general expectation for ``natural size" of  $\bar c^Q(t)$ of order unity and
 is about 20 times smaller
and has an opposite sign than the result of the bag model calculation in Ref.~\cite{Ji:1997gm}.
This smallness implies very interesting physics picture -- the quark and gluon subsystems inside the nucleon interact weakly.
The positive sign of $\bar c^Q(t)$ obtained here implies that the force acting on the quarks from side of gluons corresponds to squeezing (compression) of the quark subsystem.
We estimated
that the contribution of $\bar c^Q (t)$ to the pressure distribution inside the nucleon  is in the range of $1 -20$\%  relative to the
contribution of the quark $D$-term.

Finally we note that the smallness of $\bar c^Q(t)$ we obtained here can be directly related to Teryaev's equipartition conjecture  \cite{Teryaev:2006fk,Teryaev:2016edw}.

\noindent
\section*{\normalsize \bf Acknowledgements}

\noindent
MVP is grateful to C\'edric Lorc\'e , Peter Schweitzer and Oleg Teryaev for illuminating discussions.
This work is supported by the Sino-German CRC 110
``Symmetries and the Emergence of Structure in QCD".


\end{document}